\documentclass[printer,floatfix]{aa} 
\usepackage{graphicx}

\begin{document}
\def\al{&\!\!\!\!}
\def\x{{\bf x}}
\def\f{\frac}
\def\y{\frac{1}{2}}


\title{Tully-Fisher relation, key to dark companion of baryonic matter }

\author{ Yousef Sobouti,
         \thanks{\email{sobouti@iasbs.ac.ir}}
          Akram Hasani Zonoozi,
        \thanks{\email{a.hasani@iasbs.ac.ir}}
          \and
         Hosein Haghi
        \thanks{\email{haghi@iasbs.ac.ir}}  }

 \institute{ Institute for Advanced Studies in Basic Sciences (IASBS), P. O. Box 45195-1159,
    Zanjan, Iran }

\date{Received *****; Accepted *****}

\abstract{ Rotation curves of spiral galaxies \emph{i}) fall off
much less steeply than the Keplerian curves do, and \emph{ii}) have
asymptotic speeds almost proportional to the fourth root of the mass
of the galaxy, the Tully-Fisher relation. These features alone are
sufficient for assigning a dark companion to the galaxy in an
unambiguous way.  In regions outside a spherical system, we design a
spherically symmetric spacetime to accommodate the peculiarities
just mentioned. Gravitation emerges in excess of what the observable
matter can produce. We attribute the excess gravitation to a
hypothetical, dark, perfect fluid companion to the galaxy and resort
to the Tully-Fisher relation to deduce its density and pressure. The
dark density turns out to be proportional to the square root of the
mass of the galaxy and to fall off as $r^{-(2+\alpha)}, ~\alpha\ll
1$. The dark equation of state is barrotropic. For the interior of
the configuration, we require the continuity of the total force
field at the boundary of the system. This enables us to determine
the size and the distribution of the interior dark density and
pressure in terms of the structure of the observable matter. The
formalism is nonlocal and nonlinear, and the density and pressure of
the dark matter at any spacetime point turn out to depend on certain
integrals of the baryonic matter over all or parts of the system in
a nonlinear manner.  }
\authorrunning{Sobouti et al.}
\titlerunning{dark companion of baryonic matter}
\maketitle

\section{Introduction}
Gravitation of the observable matter in galaxies and clusters of
galaxies is not sufficient for explaining their dynamics. Dark
matter scenarios and/or alternative theories of gravitation (see e.
g., Milgrom 1983; Behar and Carmelli 2000; Capozziello et al. 2002,
2003 and 2006; Carroll et al. 2004; Norjiri et al. 2003 and 2004;
Moffat 2005; Sobouti 2007) are called in to resolve the dilemma. The
fact remains, however, that the proponents of dark matter have
always looked for it in observable matter.  No one has, so far,
reported a case where there is still no baryonic matter, but there
is a dynamical issue to be settled. In view of this negative
observation, it has been conjectured (\cite{sob}) that, if there is
a dark companion to any baryonic matter, there must be rules to
connect the properties of the twin entities.  On the other hand, the
existence of such a rule will entitle one to interpret the case as
an alternative gravity, thus reducing the difference between the two
paradigms to the level of semantics. This conclusion, however, is
true as long as the assumed dark matter does not interact with the
baryonic one in any other way than through its gravitation.

Sobouti assumes a spherically symmetric system, attributes a dark
perfect fluid companion to it, and requires the rotation curve of
the system to display the same asymptotic behavior as those of the
actual spirals. The reason for the assumption of a dark fluid
instead of the conventionally assumed dark pressureless dust, is to
ensure the satisfaction of the Bianchi identities and thereby the
baryonic conservation laws (See sect. 8 for further explanation).

In regions outside to the baryonic system, he finds the density and
pressure of the dark fluid companion in terms of the mass of the
host system. The Tully-Fisher relation and the slow non-Keplerian
decline of the rotation curves play key roles in determining the
relation between the matter and its dark twin.

In this paper, we follow the same line of argument to find the
structure of the dark matter in the interior of the baryonic system.
The continuity of the total gravitational force at the boundary of
the observable matter leads to the dark matter distribution in the
interior. The Tully-Fisher relation is a nonlocal and nonlinear
feature of the dynamics of galaxies: a) The presence of the total or
partial integrals of the baryonic matter in the structure of both
exterior and interior solutions reflects the nonlocality.  b) That
the excess gravitation does not increase proportionally upon
increasing the mass of the host galaxy indicates the nonlinearity.
To emphasize these two features, we refer to the formalism developed
here as the nonlocal and nonlinear (NN) one.

To check its validity, the formalism is applied to NGC 2903 and NGC
1560, two examples of high and low surface brightness galaxies,
respectively, and the resulting rotation curves are compared with
those obtained through other approaches.

\section{Model and formalism}
The following is a brief background from Sobouti (2008a, b; 2009).
The physical system is a spherically symmetric baryonic matter of
finite extent. By conjecture there is a  dark presence that pervades
both the interior and exterior of the system. The spacetime metric
inside and outside of the system is necessarily spherically
symmetric and takes the form
\begin{eqnarray}
ds^2 = -B(r) dt^2 + A(r) dr^2 + r^2 ( d\theta^2 + \sin^2\theta
d\varphi^2). \label{eq1}
\end{eqnarray}
Let both the baryonic matter and its dark companion be perfect
fluids of densities $\rho, ~\rho_d$, of pressures $p,~p_d$,
respectively, and be at rest. From the field equations of general
relativity (GR), we find
\begin{eqnarray}\label{eq2}
\f{1}{r^2}\left[\f{d}{dr}\left(\f{r}{A}\right)-1\right] = -(\rho +
\rho_d),
\end{eqnarray}
\begin{eqnarray}\label{eq3}
\f{1}{r A}\left(\f{B'}{B} + \f{A'}{A}\right) =
[(p+p_d)+(\rho+\rho_d)],
\end{eqnarray}
where we have let $8\pi G = c^2 = 1$, and `$'$'$=d/dr$. In the
nonrelativistic regime, we neglect the pressures, eliminate the
densities between the two equations, and arrive at
\begin{eqnarray}\label{eq4}
\f{B'}{B} = \f{1}{r}(A-1).
\end{eqnarray}
In the following two sections we solve Eqs. (\ref{eq2}) -
(\ref{eq4}) inside and outside the baryonic system.

\section{Exterior solution}
Hereafter, the parameters pertaining to the interior and exterior of
the system will be labeled  by the superscripts $(i)$ and $(e)$,
respectively. The unknowns in Eqs. (\ref{eq2}-\ref{eq4}) are $A,~ B,
~\rho_d,~ p_d,$ and the dark equation of state. We begin with Eq.
(\ref{eq4}) and assume that in the baryonic vacuum, $\rho=p=0$, the
factor $(A^{(e)}-1)$ is differentiable and has the series expansion
\begin{eqnarray}\label{eq5}
(A^{(e)}-1) = \left(\f{r_0}{r}\right)^\alpha \left( s_0+ \f{s_1}{r}+
\cdots\right),~~~r\geq R,  
\end{eqnarray}
where the indicial exponent $\alpha$ and $s_0$ are dimensionless,
$s_1$ has the dimension of length, $r_0$ is an arbitrary length
scale of the system, and $R$ is the radius of the baryonic sphere.
Substituting Eq. (\ref{eq5}) into Eq. (\ref{eq4}) and integrating
the resulting expression, gives
\begin{eqnarray}\label{eq6}
B^{(e)} = \exp\left[
-\left(\f{r_0}{r}\right)^\alpha\left(\f{s_0}{\alpha}+\f{s_1}{(1+\alpha)r}+\cdots\right)\right].
\end{eqnarray}
We expand the exponential factor, keep its first two terms, and for
the weak field gravitational potential, $\phi= (B-1)/2$, find
\begin{eqnarray}\label{eq7}
\phi^{(e)} = -\f{1}{2}\left(\f{r_0}{r}\right)^\alpha\left[
\f{s_0}{\alpha}+\f{s_1}{(1+\alpha)r}+\cdots\right].
\end{eqnarray}
The square of the circular speed of a test object orbiting the
galaxy is
\begin{eqnarray}\label{eq8}
v^2 = r \f{d \phi^{(e)}}{d r} =
\f{1}{2}\left(\f{r_0}{r}\right)^\alpha \left( s_0+ \f{s_1}{r}+
\cdots \right).
\end{eqnarray}
Equation (\ref{eq8}) is the rotation curve of our hypothetical
galaxy in its baryonic vacuum. It has an asymptotically constant
logarithmic slope,
$$\Delta = d\ln v^2/d \ln r\rightarrow-\alpha~ ~\textrm{as}~r\rightarrow
\infty.$$

\subsection{Determination of $\alpha,~s_0,~ s_1,~\cdots~$}
\noindent Rotation curves of actual spiral galaxies have two
distinct non-classical features:

\emph{i}) Their asymptotic slopes are much flatter than that of the
Keplerian curves, $-1$, (\cite{san96}; \cite{bos83}; \cite{beg89};
\cite{ps95}; \cite{bbs91}; \cite{sv98}; \cite{smcg02}). This implies
$\alpha\ll 1$. From \cite{pss96}, who study 1100 galaxies with the
aim of arriving at a universal rotation curve, we estimate
\begin{eqnarray}\label{eq9}
\alpha< 0.01.
\end{eqnarray}
Moreover, $\alpha$ does not seem to be a universal constant. The
rotation curves of more massive galaxies appear to fall off somewhat
more steeply than those of the less massive ones (\cite{pss96}).
Hereafter, for simplicity but mainly for pedagogical reasons, we
work in the limit of $\alpha\rightarrow 0$.

\emph{ii})  Their asymptotic speeds follow the Tully-Fisher
relation.  They are almost proportional to the fourth root of the
mass of the host galaxy (\cite{tf77}; \cite{beg89}; \cite{mcg00};
\cite{mcg05}). In Eq. (\ref{eq8}), letting $\alpha\rightarrow 0$,
the dominant term at large distances is $v^2= s_0/2$.  We identify
this $v$ with the Tully-Fisher asymptote and conclude that
\begin{eqnarray}\label{eq10}
s_0 = \lambda\left({M}/{M_\odot}\right)^{1/2},~~\lambda= 2.8 \times
10^{-12},
\end{eqnarray}
where M is the galactic mass, and $\lambda$ can be obtained either
from a direct examination of the observed asymptotic speeds
(\cite{sob07}) or from a comparison of the first term of Eq.
(\ref{eq8}) with the low acceleration limit of MOND (\cite{mil83}):
$ v^2/r\rightarrow (a_0 g_N )^{1/2},~ a_0= 1.2\times 10^{-10}
\textrm{m sec}^{-2}$ (\cite{beg89}).

Again letting $\alpha\rightarrow 0$, the second term in Eq.
(\ref{eq8}) is the classic Newtonian or GR term. Therefore, $s_1$
should be identified with the Schwarzschild radius of the host
galaxy:
\begin{eqnarray}\label{eq11}
s_1 = 2 GM/c^2.
\end{eqnarray}
Here, for clarity, we have restored the constants $c^2$ and $G$ and
written $s_1$ in physical units. There is no compelling
observational evidence to indicate the need for other terms in Eqs.
(\ref{eq5}) - (\ref{eq8}).  Therefore, at least at  the present
state of the extent and accuracy of the observational data, we
truncate the series at the $s_1$ term.

\section{Interior solution}
The first and foremost condition to be satisfied is the continuity
of the total force exerted on a test object at the boundary, $R$, of
the baryonic system.  Pressure forces are anticipated to be
insignificant in the present problem so are ignored. The
gravitational forces remain. From Eqs. (\ref{eq8}) - (\ref{eq11}),
the exterior force is
\begin{eqnarray}\label{eq12}
\f{d\phi^{(e)}}{d r} =
\f{1}{2}\left[\lambda\left(\f{M}{M_\odot}\right)^{1/2}\f{1}{r} +
\f{2GM}{c^2}\f{1}{r^2}\right],~r\geq R.
\end{eqnarray}
By analogy, for the interior of the system we adopt
\begin{eqnarray}\label{eq13}
\f{d\phi^{(i)}}{d r} =
\f{1}{2}\left[\lambda\left(\f{M(r)}{M_\odot}\right)^{1/2}\f{1}{r} +
\f{2GM(r)}{c^2}\f{1}{r^2}\right],~r\leq R,
\end{eqnarray}
where $M(r)=4\pi\int_0^r \rho r^2 d r$ is the variable baryonic mass
inside the radius $r$. The continuity of the exterior and interior
forces at the boundary is evident, QED. Once the baryonic $\rho(r)$
and $M(r)$ are known, $\phi^{(i)}(r)$, $B^{i}(r) \approx 1 +
2\phi^{(i)}$  and $A^{(i)}$ can be integrated.  The expression for
the latter is much simpler and is given below for later reference.
From Eq. (\ref{eq4}) we find
\begin{eqnarray}\label{eq14}
A^{(i)} - 1 = 2 r \f{d\phi^{(i)}}{d r} =
\left[\lambda\left(\f{M(r)}{M_\odot}\right)^{1/2} +
\f{2GM(r)}{c^2}\f{1}{r}\right].
\end{eqnarray}
This has the same form as Eq. (\ref{eq5}), where $M$ is replaced by
$M(r)$.

\section{Structure of the dark matter}
The densities are obtained from Eq. (\ref{eq2}) or equivalently
from Poisson's equation through Eqs. (\ref{eq12}) - (\ref{eq13}).
For the exterior dark density we find
\begin{eqnarray}\label{eq15}
\rho_d^{(e)}(r)=
\lambda\left(\f{M}{M_\odot}\right)^{1/2}\f{1}{r^2},~~~r\geq R.
\end{eqnarray}
Note the square root dependence on the mass of the galaxy and the
fall out as $r^{-2}$. For the interior, $A^{(i)}$ is given by Eq.
(\ref{eq14}), whose first term gives the interior dark density and
the second renders the baryonic density, $\rho$. Thus,
\begin{eqnarray}\label{eq16}
\rho_d^{(i)}(r) =
\lambda\left[\f{M(r)}{M_\odot}\right]^{1/2}\f{1}{r^2}\left[ 1+
2\pi \f{\rho r^3}{M(r)}\right],~~r\leq R.
\end{eqnarray}
The dark matter inside the radius $r$
is
\begin{eqnarray}\label{eq17}
M_d(r) = 4\pi\int_0^r \rho_d(r) r^2 d r= \lambda
\left[\f{M(r)}{M_\odot}\right]^{1/2} r. 
\end{eqnarray}
Equation (\ref{eq17}) holds for any $r$. For $r\geq R$, however,
$M(r)$ attains its maximum constant value, $M$,
and $M_d(r>R)$ becomes proportional to $r$.

 It is instructive to look at the behavior of Eq. (\ref{eq16}) in
the neighborhood of the origin, where $\rho\rightarrow\rho_c$ and $
M(r) \rightarrow 4\pi \rho_c r^3/3$.  Equation (\ref{eq16}) tends
toward
\begin{eqnarray}\label{eq20}
\rho_d^{(i)}(r\rightarrow 0)=
\f{5}{2}\lambda\left(\f{4\pi}{3}\f{\rho_c}{M\odot}\right)^{1/2}r^{-1/2}.
\end{eqnarray}
Similarly,
\begin{eqnarray}\label{eq21}
M_d^{(i)}(r\rightarrow 0)=
\lambda\left(\f{4\pi}{3}\f{\rho_c}{M\odot}\right)^{1/2}r^{5/2}.
\end{eqnarray}
While the density becomes singular as $r\rightarrow 0$, no cusp
develops. For the measure $r^2 dr$ tends to zero as $r\rightarrow
0$.

Pressures of the matter and of its dark companion are obtained
from their hydrostatic equilibrium, a requirement of the Bianchi
identities. The general formula is
\begin{eqnarray}\label{eq22}
\f{p'}{p+\rho} \approx \f{p'}{\rho}= -\f{1}{2}\f{B'}{B}\approx
-\f{d\phi}{dr}.
\end{eqnarray}
For the exterior pressure from Eqs. (\ref{eq22}), (\ref{eq15}),
(\ref{eq12}), we find
\begin{eqnarray}\label{eq23}
p_d^{(e)}(r) =\f{1}{4} s_0 \left( \f{s_0}{r^2} +
\f{2}{3}\f{s_1}{r^3}\right),~~r \geq R.
\end{eqnarray}
The presence of an extra factor of $s_0$ in Eq.(\ref{eq23}) makes
the pressure an order of magnitude less than the density and
justifies the approximation made in the derivation of Eq.(\ref{eq4})
and thereafter. The equation of state, $p(\rho)$, in the exterior
region is obtained by eliminating $r$ between Eqs. (\ref{eq23}) and
(\ref{eq15}). It is barrotropic.  The internal pressure is obtained
in a similar way.  It is, however, an involved expression so is too
involved expression to give here.

A pedagogical note: Throughout the text, except in Eq.(\ref{eq11}), we have chosen
$8 \pi G = c^2 = 1$. To write the results in physical units, the rule
is to multiply, everywhere, the potentials, $\phi$, by $c^2$,
the dark densities, $\rho_d$, dark masses, $M_d$, by $c^2/8\pi G$,
and the dark pressures, $p_d$, by $c^4/8\pi G$.

\section{ Application to actual spirals}
Spiral galaxies are flattened objects. Their  approximation as
spherical systems introduces an error on the order of
$(R_{gyr}/r)^2$, where  $R_{gyr}(r)$ is the gyration radius of the
mass enclosed within a radius $r$. In a flat system that thins out
as an exponential or as a Kuzmin disk, say, this ratio would be a
few parts in thousand and small enough for our purpose. This is also
the practice of all the authors quoted so far in this paper. To
illustrate the practical applicability of the formalism developed
here, we construct the rotation  curves of two standard high- and
low- surface brightness galaxies and compare the results with those
obtained through MOND's formalism.

NGC 2903 is a textbook example of a high surface brightness spiral.
It has a large stellar component and small HI content. The gas is
confined to the galactic plane and follows circular orbits. It is
well observed out to about 40 kpc (\cite{beg89}).  In contrast, NGC
1560 is a low surface brightness spiral with a dominant gas
component.  Its observed rotation curve extends out to about 8 kpc
and does not seem to have reached its asymptotic regime.

In Fig. 1 we construct the rotation curves of our NN formalism from
Eq. (\ref{eq13}), in which $M(r)$ is the total, stellar plus HI,
mass interior to $r$. The free adjustable parameter in matching the
theoretical curves to data points, is the `stellar' mass-to-light
ratio, $\Upsilon$, assumed to be constant throughout the galaxy. For
comparison we have also included the rotation curves of MOND. That
the NN curves trace the data points more closely than the MOND ones
can be seen pictorially.
The $\chi^2$ test and $\Upsilon$'s of Table 1, however, illustrate
this in a quantitative way. In both galaxies our $\chi_{NN}^2$ is
noticeably small. Significant, however, is the low stellar
mass-to-light ratio of the young and gas-dominated NGC 1560. Our
$\Upsilon=0.3$ is, by far, closer to 0.4 estimate of McGaugh (2002)
than to 1.1 of MOND.
\begin{figure}{}
\begin{center}
\resizebox{10cm}{!}{\includegraphics[width=70mm,height=100mm]{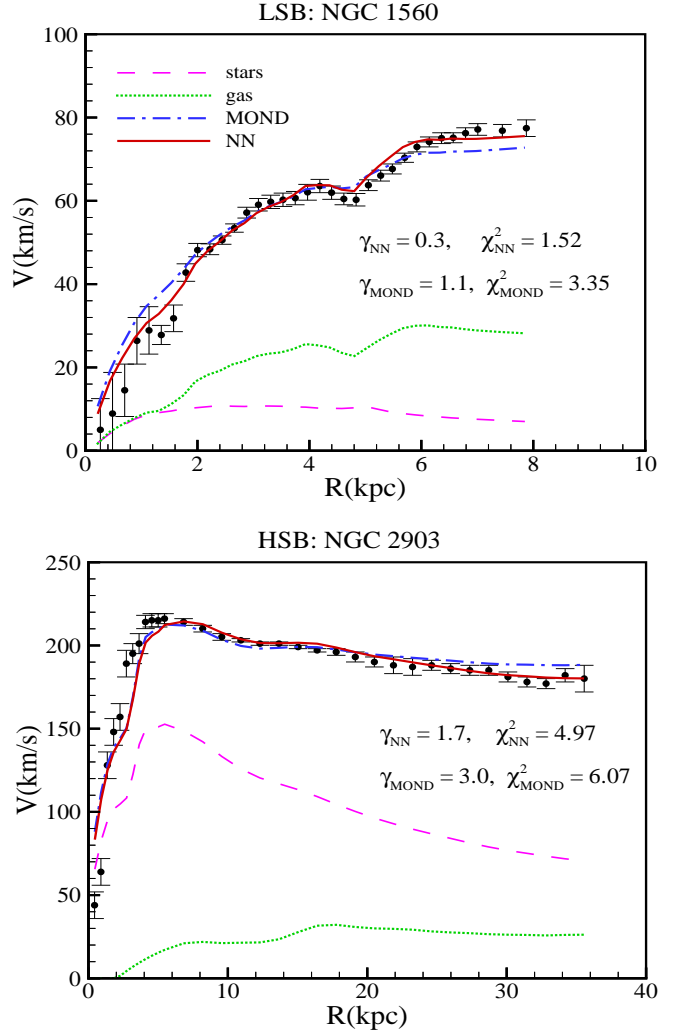}}
\caption {Points with error bars are observed data. Dotted and
dashed lines are the contributions of the gaseous and stellar
components to the rotation curves, respectively. Dashed-dotted line
is the rotation curve constructed through MOND's formalism.  Solid
line is our rotation curve calculated from Eq. (\ref{eq13}). The
free parameter in matching theoretical curves to data points, is the
stellar mass-to-light ratio.}\label{fig1}
\end{center}
\end{figure}

\begin{table}
\caption{Minimum  $\chi^2$ and fitted stellar mass-to-light ratio,
$\Upsilon$, of MOND and of our NN formalism. \label{table}}
\begin{tabular}{cc  c  c  c  c }\\
Galaxy && $\chi^{2}_{NN}$ &$\Upsilon_{NN}$& $\chi^{2}_{MOND}$  & $\Upsilon_{MOND}$ \\
\hline~~~~~~\\
NGC 2903&& 4.97 & 1.7 & 6.07 & 3.0 \\
NGC 1560&& 1.52 & 0.3 & 3.35 & 1.1 \\
\end{tabular}
\end{table}

Our next project is to study pressure-supported systems, globular
clusters and dwarf spheroidal galaxies (dSph). Globular clusters are
commonly believed to be almost Newtonian systems, while dSph's show
significant deviations from Newtonian regimes.  The low baryonic
mass and extremely high dynamical-mass-to-light ratio of dSph's are
inconsistent with population synthesis models (\cite{hilk06};
\cite{jor09}; \cite{ang08}). Our approach is to find a counterpart
of the classical virial theorem for our proposed gravity and to
solve a modified Jeans equation. The aim is to verify whether the
velocity dispersions obtained via Jeans equation fit the observed
data. We also hope to be able to come up with a notion equivalent to
the fundamental plane for galaxies where one arranges the galaxies
on a two-parameter-plane in a three-dimensional space of luminosity,
velocity dispersion, and some other global characteristics of the
galaxies.

\section{Nonlocality and nonlinearity of the formalism:}
The masses $M$ and $M(r)$ are integrals over all or parts of the
system. Their presence, in the structure of the spacetime metric, in
the rotation curve, in the expressions for the dark densities and
pressures, etc., reflects the nonlocal nature of the theory. That
these integrals enter the formalism not in a linear way indicates
the nonlinearity of it. Both features are rooted in the Tully-Fisher
relation, which requires the dynamical variables at one spacetime
point to depend on the integral properties of the whole or parts of
the system through the square root of these integrals. Any attempt
to derive the spacetime metric entertained in this paper through a
variational principle should take these two features into account.

In this respect, Hehl and Mashhoon's generalization of GR,
(\cite{hm09ab}), constructed within the framework of the
translational gauge theory of gravity, is interesting.  In the weak
field approximation, the excess gravitation coming from the
nonlocality of their theory can be interpreted as a dark companion
to the baryonic matter. In the case of a point baryonic mass, $M$,
the dark density has the expected $r^{-2}$ distribution, But it does
not obey the Tully - Fisher relation.
Instead of $M^{1/2}$, it is proportional to $M$ itself.\\

\section{Concluding remarks}
The  formalism developed here is a dark matter scenario or,
equivalently, a modified GR paradigm to understand the non-classical
behavior of the rotation curves of spiral galaxies. Following
(\cite{sob}), we attribute a hypothetical dark perfect fluid
companion to our model galaxy , and find the size and the
distribution of the companion by comparing the rotation curve of the
model with those of the actual galaxies. However, as long as the
dark companion displays no physical characteristics other than its
gravitation, one has the option to interpret the scenario as an
alternative theory of gravitation. Here, for example, one may
maintain that the gravitation of a baryonic sphere is not what
Newton or Schwarzschild profess, but rather what one infers from the
spacetime metric detailed above.  In fact we wish to emphasize that
any modified gravity is expressible in terms of a dark matter
scenario. And vice versa, any dark matter paradigm, in which the
matter and its dark twin are related by certain rules, is
explainable by a modified gravity.  The difference between the two
alternatives is semantic.

Dynamics of galaxies is a nonrelativistic issue. Yet, its analysis
in a GR context answers questions that otherwise are left out. In
particular, in a nonrelativistic scenario, there is neither need nor
logic to assign a pressure field to a hypothetical matter that one
knows nothing about its nature. In a GR context, on the other hand,
the dark matter has to have a pressure field and has to be in
hydrodynamic equilibrium as a requirement of the Bianchi identities
and thereby of the conservation laws of the baryonic matter, i.e.
the vanishing of the 4-divergence of both sides of the field
equations. Let us also note in passing that all those metric
approaches that attempt to explain the galaxy problems with the aid
of a single scalar field are subject to the same criticism, namely
the violation of the Bianchi identities and of the conservation
laws.

Regions outside to the baryonic matter are not dark matter vacua.
Therefore, the Ricci scalar does not vanish, and there are excess
lensing and excess periastron precession caused by the dark matter.
These are analyzed in Sobouti (2008a, b; 2009).

The formalism is good for spherical distributions of baryonic
matters.  An axiomatic generalization to nonspherical configurations
or to many body systems requires further deliberations and more
accurate observational data to help find some solutions. One might
need other postulates not contemplated. The difficulty lies in the
nonlinearity of the formalism. There is no superposition principle.
One may not add the fields of the dark companions of two separate
baryonic systems because $s_0$ of Eq. (\ref {eq10}) is not linear in
$M$ or $M(r)$.

\begin{acknowledgements}
We thank S. S. McGaugh for providing us with observational data on
the rotation of  galaxies and for his useful comments on their
interpretation.
\end{acknowledgements}

\end{document}